# Oxygen adsorption induced superconductivity in ultrathin FeTe film on SrTiO$_3$(001)

Wei Ren[1,3], Hao Ru[2], Kun Peng[3], Huifang Li[3], Shuai Lu[3], Aixi Chen[3], Pengdong Wang[3], Xinwei Fang[3], Zhiyun Li[3], Rong Huang[3], Li Wang[3,*], Yihua Wang[2,4,*], Fangsen Li[1,3,*]

[1] School of Nano-Tech and Nano-Bionics, University of Science and Technology of China, Hefei, 230026, China

[2] Department of Physics and State Key Laboratory of Surface Physics, Fudan University, Shanghai China 200438

[3] Vacuum Interconnected Nanotech Workstation (Nano-X), Suzhou Institute of Nano-Tech and Nano-Bionics (SINANO), Chinese Academy of Sciences (CAS), Suzhou 215123, China

[4] Shanghai Research Center for Quantum Sciences, Shanghai 201315, China

*Corresponding authors: E-mail: fsli2015@sinano.ac.cn (F. Li), lwang2017@sinano.ac.cn (L. Wang) and wangyhv@fudan.edu.cn (Y. Wang)

## Abstract

The phenomenon of oxygen incorporation induced superconductivity in iron telluride (Fe$_{1+y}$Te, with antiferromagnetic (AFM) orders) is intriguing and quite different from the case of FeSe. Until now, the microscopic origin of the induced superconductivity and the role of oxygen are far from clear. Here, by combining *in-situ* scanning tunneling microscopy/spectroscopy (STM/STS) and x-ray photoemission spectroscopy (XPS) on oxygenated FeTe, we found physically adsorbed O$_2$ molecules crystallized into *c*(2/3×2) structure as an oxygen overlayer at low temperature, which was vital for superconductivity. The O$_2$ overlayer were not epitaxial on the FeTe lattice, which implied weak O$_2$-FeTe interaction but strong molecular interactions. Energy shift observed in the STS and XPS measurements indicated hole doping effect from the O$_2$ overlayer to the FeTe layer, leading to a superconducting gap of 4.5 meV opened across the Fermi level. Our direct microscopic probe clarified the role of oxygen on FeTe and emphasized the importance of charge transfer effect to induce superconductivity in iron-chalcogenide thin films.

# 1. Introduction

Iron-chalcogenide Fe(Se,Te) superconductors are an important family of ($T_c$) iron-based high transition temperature ($T_c$) superconductors. The monolayer grown on SrTiO$_3$ (STO) still holds the record in $T_c$ with much-enhanced superconducting pairing [1-3] ; the strong spin-orbit coupling in Te-alloyed compounds leads to a topological band structure [4-7] , the family also exhibits rich magnetic phases strongly dependent on the chemical composition [8, 9]. Non-superconducting FeTe with antiferromagnetic (AFM) orders was considered as the parent compound of Iron-chalcogenide superconductors [10]. Superconductivity could be induced in FeTe through Se or S substitution [10] and oxygen incorporation by low-temperature annealing or long-time exposure in an O$_2$ atmosphere [11-21]. Despite lots efforts towards understanding of oxygenated FeTe bulk crystals [15-22] and thin films [11-14, 21], the microscopic role played by oxygen is still unknown.

Zheng et al. [12] found that interstitial oxygen, rather than oxygen substitution [23], was responsible for the emergence of superconductivity. The interstitial oxygen [13, 16] would dope hole into FeTe [17], suppress the AFM ordering and then induce superconductivity. Yamazaki et al. [19] and Sun et al. [18] have studied the dynamics of oxygen annealing and concluded that the superconductivity first emerged on the surface and the superconducting regions moved inside with non-superconducting materials left on the surface, such as Fe$_2$O$_3$, TeO$_x$, FeTe$_2$. The controversial results from different experimental probes call for a unified study of the crystal, chemical and electronic structures on thin films with controlled oxygen incorporation. However, this is generally challenging to achieve for *ex-situ* measurements because samples exposed to atmosphere inevitably changes in surface morphology and chemistry. In this study, we overcame this challenge by integrating in one ultra-high vacuum environment for the sample growth and oxygenation process along with spectroscopic tools with elemental and spatial resolution.

Firstly, we grew nearly stoichiometric 10 Unit-Cell (UC) FeTe epitaxy ultrathin films by molecular beam epitaxy (MBE), and then studied the adsorption of O$_2$ incorporation by combined *in-situ* STM and X-ray photoelectron spectroscopy (XPS). We found that

$O_2$ molecules physically adsorbed on FeTe surface under low $O_2$ partial pressure at room temperature and crystallized into *c*(2/3×2) structure, which gradually disappeared after heating up to 100℃ in vacuum. No surface oxidation can be found. Band shift observed in STS and XPS suggest hole doping from $O_2$ overlayer to FeTe, which induced a superconducting gap in FeTe. It agrees with the *ex-situ* transport measurement that long-time exposure to air can induce superconducting transition in 10UC FeTe on STO.

## 2. Materials Synthesis and Characterization

High quality 10UC FeTe thin films were grown on $SrTiO_3$(001) by using a Unisoku ultrahigh vacuum (UHV) low temperature STM system (Unisoku 1300 9T-2T-2T), equipped with an MBE chamber for film preparation. The Nb doped-STO substrate (Nb:0.7% wt, KMT comp.) was firstly annealed at 1000℃ for 1 h, and then kept at 280°C during FeTe growth. High purity Fe (99.9999%) and Te (99.9999%) sources were co-evaporated onto $SrTiO_3$(001) substrates from two standard Knudsen cells. To improve the quality of crystallization, FeTe films were annealed at 280°C for 1 hour after growth. The oxygen absorption was carried out at room temperature under $O_2$ partial pressure of $1.6\times10^{-4}$ mbar (if not specified otherwise). After $O_2$ absorption, the samples were *in-situ* transferred to STM chamber or XPS equipment for further characterizations. For electronic transport measurement, 10UC FeTe films were grown on high-resistance STO and covered with Te protection layer.

All STM/STS measurements were performed at 4.7 K with bias voltage applied to the sample. The differential conductance d*I*/d*V* spectra were acquired by using a standard lock-in technique at a frequency of 983 Hz. The STM topographic images were processed with WSXM software [24].

After STM/STS measurements, the samples were transferred *via* UHV interconnection pipeline in Nano-X for XPS/UPS characterization without exposure to air. XPS/UPS experiments were carried out by using a commercial analyzer (PHI 5000, VersaProbe, ULVAC-PHI, $\sim5\times10^{-10}$ mbar), equipped with a monochromatic Al $K_\alpha$ X-ray source of 1486.7 eV. The binding energy (BE) of core-level peaks were calibrated with respect to the C-C 1s bond (BE = 284.8 eV). A monochromatized He I with energy of

21.22 eV were used for UPS measurement. During the UPS measurement, a voltage of -5 V was applied between the sample and the spectrometer.

## 3. Results and analysis

High-quality FeTe films are epitaxially grown on STO substrates *via* step flow growth mode. Fig.1a shows the typical surface morphology of 10UC FeTe film and Fig.1b shows the atomically resolved image with Te termination. Fast Fourier transform image yields enlarged lattice constant of ~0.384 nm, similar to the case of FeSe on STO [1]. No Fe adatoms (or interstitial Fe atoms) and Te vacancies can be observed on the surface, which indicates nearly stoichiometric FeTe epitaxy films. To carry out electronic transport measurement, protection Te layer was deposited on 10UC FeTe surface, and the results were shown in Fig. 1c. No superconducting transition but semiconducting behavior was observed in initially grown 10UC FeTe film down to 1.8 K, consistent well with previous experiments [25, 26]. Compared with bulk or thick FeTe films [11, 13, 23, 27-29], we have not found anomaly in the *R-T* curve between 40 K~80 K, which further demonstrates free of Fe interstitial atoms [26]. AFM state may be suppressed here. After exposure the sample to air for 8 days, the resistance turned downward at ~ 3.2 K, as shown in the inset of Fig. 1c. Considering the fact of long-time exposure induced superconductivity in FeTe compounds, such downward should be attributed to superconducting transition. It's further verified by the suppression of transition under external magnetic fields in Fig. 1d, characteristic of superconductivity. Non-zero resistance here suggests the developed superconductivity is not uniform. We found that exposure to air for longer time would increase the superconducting transition temperature further.

To figure out the superconducting mechanism in FeTe after exposure to $O_2$, we carried out oxygenated process under a well-control condition and probed the atomic and electronic structures microscopically. Fig.2a shows the surface morphology of initial adsorption stage after exposure 10UC FeTe to $O_2$ under partial pressure of $1.6×10^{-4}$ mbar for 15 mins. We can observe extra layers along the step edge, while other uncovered clean FeTe areas still remain unchanged, as illustrated by the atomically resolved STM image in the inset. No $O_{Te}$ substitution or interstitial O atoms can be observed. The enlarged image in Fig.2b shows there are small compacted islands with a height of ~0.18 nm in the overlayer, as shown in Fig. 2c. Some dispersed clusters can be also found at the domain boundaries, which should be the ones that haven't coarsened into the islands. If further increasing the amount of inlet $O_2$, nearly the whole surface can be covered by the adsorbed overlayers, as shown in Fig. S1a. The uncovered areas are still very clean in Fig. S1b.

To unravel the nature of adsorbed overlayer ($O_2$ molecules, $FeO_x$, $TeO_x$, Fe-Te-O or others?), Fig. 2d and Fig. 2e present the core-level XPS spectra of Te 3d and Fe 2p before and after inlet $O_2$, respectively. After inlet $O_2$ under partial pressure of $1.6\times10^{-4}$ mbar, we haven't observed Te-O and Fe-O spectra signals. If exposure to $O_2$ partial pressure of $1.5\times10^{-1}$ mbar for 20 mins, as shown in Fig. S3c, clear Te-O core level peak can be identified, which suggests surface oxidation under high $O_2$ partial pressure. The XPS spectra of Te 3d and Fe 2p show obvious peak shift toward higher binding energy, which hints hole doping effect in FeTe. The overlayer would gradually desorb when heating up to 100℃, and disappeared after annealing at 200℃ for 2 hours. Only clean FeTe surface was left. The overlayer will re-adsorb after inlet $O_2$ again. Thus, we can rule out the possibility of Fe- or Te-related compounds, because there was negligible residual Fe or Te atoms after growth. In addition, we notice there are some dispersed clusters higher than the compacted islands in Fig. 2b and Fig. 3a, which helps us to conclude that the overlayer should be $O_2$ molecules rather than stacked O atoms. The shape of O 1s core level almost remain unchanged after $O_2$ adsorption under partial pressure of $1.6\times10^{-4}$ mbar, as illustrated in Fig. S3, but change a lot after surface oxidation, due to the signal from O-Te and O-Fe. It implies the signal of O 1s at 529.5 eV ~ 531 eV are mainly from the STO substrate, which as expected exhibits similar core level shift as Te 3d, as shown in Fig. S3a. It also consists with the nature of physically adsorbed $O_2$ molecules, where characteristic core-level XPS peak always can't be found. Thus, we name the overlayer heterostructure as FeTe:$O_2$. UPS measurements in Fig.2f show that the work function just slightly changed from $\phi_{FeTe}$ ~ 4.43 eV to $\phi_{FeTe:O2}$ ~ 4.38 eV after $O_2$ adsorption. The small difference in work function is also consistent with the nature of $O_2$ molecules in the overlayer.

Fig. 3a shows some small compacted FeTe:$O_2$ islands, where striped structure can be clearly observed. Only two perpendicular orientations were distinguished, which consist with the tetragonal symmetry of underlying FeTe lattice. Atomically resolved structure of adsorbed FeTe:$O_2$ islands is shown in Fig. 3b. The distance between neighboring bright rows is ~0.386 nm, nearly equal to the lattice constant $a_{FeTe}$ of FeTe. The corresponding fast Fourier transform image in Fig.3c yields in-plane lattice of $n_1$ with value of 0.25±0.02 nm and $n_2$ with value of 0.41±0.02 nm. The angle between $n_1$ and $n_2$ is measured to be ~72º. We notice that the spot distance of 0.25 nm along $n_1$ is nearly equal to $2/3 a_{FeTe}$. Thus, the lattice can be notated as $c(2/3×2)$, as shown in Fig.3b and Fig.3c. Such lattice arrangement demonstrates a close relationship between $O_2$ adsorption overlayer (FeTe:$O_2$) and FeTe surface. To locate the precise adsorption sites of $O_2$ on FeTe, we achieved atomically resolved image on both the FeTe:$O_2$ island and FeTe surface in Fig.3d. The results are shown in the structural model in Fig.3e. As illustrated, along $n_1$ direction the adsorption sites of $O_2$ molecules are not well-defined relative to FeTe lattice and not at high-symmetry points. It not only signifies that the van der Waals interaction between $O_2$ molecules plays a vital role in the overlayer, but also verifies the absence of chemical bonding between the $O_2$ interlayer and the FeTe surface.

To explore how the electronic structure changes due to $O_2$ absorption, we conducted detailed differential conductance d$I$/d$V$ measurements, which are known to be proportional to local density of state (LDOS). On as-prepared and uncovered clean 10UC FeTe surfaces, the STS remain nearly unchanged with "V" shape across the Fermi level in Fig.4a, consistent with previous studies [2, 30]. A hump at negative bias voltage shift from -0.37 V to -0.325 V on FeTe:$O_2$ after $O_2$ adsorption, which indicates hole doping effect to FeTe, agreeing well with the observation of peak shift in XPS experiments in Fig.2. Such oxygen induced hole doping effect [31] was recently observed on the $CsV_3Sb_5$ with increased superconducting transition temperature. And hole doping effects were observed in $Bi_2Te_3$/FeTe [32, 33] and $Sb_2Te_3$/FeTe [26] heterostructures, and thought to be the origin of two-dimensional superconductivity in the heterostructures. Fig.4b shows the d$I$/d$V$ spectra in small energy range. Clear V-shape was observed on clean FeTe surface, similar to the one on the monolayer FeTe [2] and thick FeTe film [30]. Strikingly, on FeTe:$O_2$ islands two pronounced peaks with an energy gap of ~ 9 meV were clearly identified across the Fermi level. Considering the facts of previous observation of hole doping induced superconductivity in $Bi_2Te_3$/FeTe heterostructure and the superconductivity observed in 10UC FeTe/STO sample after long-time exposure to air, we conclude that the gap on FeTe:$O_2$ is just superconducting gap with $\Delta$ ~ 4.5 meV, larger than the one on $Bi_2Te_3$/FeTe heterostructure (~2.5 meV) [32, 34]. It should be due to interfacial effect on STO substrate, because similar enhanced superconductivities have been observed on the Pb islands [35] and monolayer $FeTe_xSe_{1-x}$ [1, 2, 25]. We have studied the spatial distribution of superconducting gap on the FeTe:$O_2$ islands in Fig.4c. Pronounced coherent peaks can be observed in all the d$I$/d$V$ curves.

The observation of enhanced superconducting gap on FeTe:$O_2$ islands is intriguing and distinct from all the previous proposed mechanism. Physically adsorbed $O_2$ molecules rather than interstitial oxygen or oxygen substitution is responsible for superconductivity in oxygenated FeTe. Previously, annealing under elevated temperature or quite long-time aging may be aimed at formation of a uniform $O_2$ overlayer into FeTe layer. Our study here also emphasizes the importance of hole doping effect and interfacial effect to induce superconductivity on FeTe. We haven't observed superconducting gap on FeTe:$O_2$ overlayer on monolayer FeTe, which may be due to the extra electron doping from STO substrate.

## 4. Conclusions

In summary, we have systematically investigated the changes of morphology and electronic structure of 10UC FeTe films after $O_2$ adsorption. E*x-situ* transport measurements demonstrate long-term exposure to air could induce superconducting transition in 10UC FeTe ultrathin film, which verifies the effect of oxygen incorporation. Initially $O_2$ molecules physically adsorbed on FeTe surface and formed compacted FeTe:$O_2$ islands with c(2/3×2) structure at low temperature. No oxygen related defects can be detected on the uncovered clean FeTe surface. Energy shifts in STS and XPS measurements point out obvious hole doping into FeTe from oxygen overlayer. Such hole doping results into a superconducting gap of ~ 4.5 meV opened with apparent coherent peaks. Our study here directly provides microscopical insights about the role of oxygen on FeTe thin film.

**Supplementary Materials:** The following are available online at XXX.

**Author contributions**

**Funding:**

This work is supported by National Natural Science Foundation of China (Grants No. 11604366, No. 11634007). F. Li acknowledges support from the Youth Innovation Promotion Association of Chinese Academy of Sciences (2017370). Y. Wang would like to acknowledge partial support by the Ministry of Science and Technology of China under Grant numbers 2017YFA0303000 and 2016YFA0301002, NSFC Grant No. 11827805, and Shanghai Municipal Science and Technology Major Project Grant No.2019SHZDZX01.

**Institutional Review Board Statement:** Not applicable.

**Informed Consent Statement:** Not applicable.

**Data Availability Statement:**

The data presented in this study are available on request from the corresponding authors.

**Acknowledgments:**

We thank Xu-cun Ma, Li-li Wang and Haiping Lin for helpful discussion.

**Conflict of interest**

The authors declare that they have no conflict of interest.

# References

1. Wang, Q.-Y.; Li, Z.; Zhang, W.-H.; Zhang, Z.-C.; Zhang, J.-S.; Li, W.; Ding, H.; Ou, Y.-B.; Deng, P.; Chang, K.; Wen, J.; Song, C.-L.; He, K.; Jia, J.-F.; Ji, S.-H.; Wang, Y.-Y.; Wang, L.-L.; Chen, X.; Ma, X.-C.; Xue, Q.-K. Interface-induced high-temperature superconductivity in single unit-cell FeSe films on $SrTiO_3$. *Chin. Phys. Lett.* **2012**, 29, 037402.

2. Li, F.; Ding, H.; Tang, C.; Peng, J.; Zhang, Q.; Zhang, W.; Zhou, G.; Zhang, D.; Song, C.-L.; He, K.; Ji, S.; Chen, X.; Gu, L.; Wang, L.; Ma, X.-C.; Xue, Q.-K. Interface-enhanced high-temperature superconductivity in single-unit-cell $FeTe_{1−x}Se_x$ films on $SrTiO_3$. *Phys. Rev. B* **2015**, 91, 220503(R).

3. Lee, J. J.; Schmitt, F. T.; Moore, R. G.; Johnston, S.; Cui, Y. T.; Li, W.; Yi, M.; Liu, Z. K.; Hashimoto, M.; Zhang, Y.; Lu, D. H.; Devereaux, T. P.; Lee, D. H.; Shen, Z. X. Interfacial mode coupling as the origin of the enhancement of T(c) in FeSe films on $SrTiO_3$. *Nature* **2014**, 515, 245-8.

4. Wang, D.; Kong, L.; Fan, P.; Chen, H.; Zhu, S.; Liu, W.; Cao, L.; Sun, Y.; Du, S.; Schneeloch, J.; Zhong, R.; Gu, G.; Fu, L.; Ding, H.; Gao, H. J. Evidence for Majorana bound states in an iron-based superconductor. *Science* **2018**, 362, 333-335.

5. Zhang, P.; Yaji, K.; Hashimoto, T.; Ota, Y.; Kondo, T.; Okazaki, K.; Wang, Z. J.; Wen, J. S.; Gu, G. D.; Ding, H.; Shin, S. Observation of topological superconductivity on the surface of an iron-based superconductor. *Science* **2018**, 360, 182.

6. Xu, G.; Lian, B.; Tang, P.; Qi, X. L.; Zhang, S. C. Topological Superconductivity on the surface of Fe-based superconductors. *Phys. Rev. Lett.* **2016**, 117, 047001.

7. Jiang, D.; Pan, Y.; Wang, S.; Lin, Y.; Holland, C. M.; Kirtley, J. R.; Chen, X.; Zhao, J.; Chen, L.; Yin, S.; Wang, Y. Observation of robust edge superconductivity in Fe(Se,Te) under strong magnetic perturbation. *Sci. Bull.* **2021**, 66, 425-432.

8. Enayat, M.; Sun, Z. X.; Singh, U. R.; Aluru, R.; Schmaus, S.; Yaresko, A.; Liu, Y.; Lin, C. T.; Tsurkan, V.; Loidl, A.; Deisenhofer, J.; Wahl, P. Real-space imaging of the atomic-scale magnetic structure of $Fe_{1+y}Te$. *Science* **2014**, 345, 653-656.

9. Trainer, C.; Yim, C. M.; Heil, C.; Giustino, F.; Croitori, D.; Tsurkan, V.; Loidl, A.; Rodriguez, E. E.; Stock, C.; Wahl, P. Manipulating surface magnetic order in iron telluride. *Sci. Adv.* **2019**, 5, eaav3478.

10. Dagotto, E. Colloquium: The unexpected properties of alkali metal iron selenide superconductors. *Rev. Mod. Phys.* **2013**, 85, 849-867.

11. Nie, Y. F.; Telesca, D.; Budnick, J. I.; Sinkovic, B.; Wells, B. O. Superconductivity induced in iron telluride films by low-temperature oxygen incorporation. *Phys. Rev. B* **2010**, 82, 020508(R).

12. Zheng M.; H. H.; Zhang C.; Mulcahy B.; Zuo J. and Eckstein J. growth and oxygen Doping of thin film FeTe by Molecular beam epitaxy. *arXiv: 1301.4696v2* **2013**.

13. Hu, H.; Kwon, J.-H.; Zheng, M.; Zhang, C.; Greene, L. H.; Eckstein, J. N.; Zuo, J.-M. Impact of interstitial oxygen on the electronic and magnetic structure in superconducting $Fe_{1+y}TeO_x$ thin films. *Phys. Rev. B* **2014**, 90, 180504(R).

14. Zhao, P. H.; Zhu, H. F.; Tian, Y. J.; Li, D. L.; Ma, L.; Suo, H. L.; Nie, J. C. O2 Annealing Induced Superconductivity in $FeTe_{1−x}Se_x$: on the Origin of Superconductivity in FeTe Films. *J. Super. Nov. Magn.* **2016**, 30, 871-876.

15. Mizuguchi, Y.; Deguchi, K.; Tsuda, S.; Yamaguchi, T.; Takano, Y. Evolution of superconductivity by oxygen annealing in $FeTe_{0.8}S_{0.2}$. *Europhys. Lett.* **2010**, 90.

16. Hu, H.; Zuo, J.-M.; Zheng, M.; Eckstein, J. N.; Park, W. K.; Greene, L. H.; Wen, J.; Xu, Z.; Lin, Z.; Li, Q.; Gu, G. Structure of the oxygen-annealed chalcogenide superconductor $Fe_{1.08}Te_{0.55}Se_{0.45}O_x$. *Phys. Rev.*

*B* **2012**, 85, 064504.

17. Su, T. S.; Yin, Y. W.; Teng, M. L.; Gong, Z. Z.; Zhang, M. J.; Li, X. G. Effect of carrier density and valence states on superconductivity of oxygen annealed $Fe_{1.06}Te_{0.6}Se_{0.4}$ single crystals. *J. Appl. Phys.* **2013**, 114.

18. Sun, Y.; Tsuchiya, Y.; Taen, T.; Yamada, T.; Pyon, S.; Sugimoto, A.; Ekino, T.; Shi, Z.; Tamegai, T. Dynamics and mechanism of oxygen annealing in $Fe_{1+y}Te_{0.6}Se_{0.4}$ single crystal. *Sci. Rep.* **2014**, 4, 4585.

19. Yamazaki, T.; Sakurai, T.; Yaguchi, H. Size Dependence of oxygen-annealing effects on superconductivity of $Fe_{1+y}Te_{1-x}S_x$. *J. Phys. Soc. Jpn.* **2016**, 85, 114712.

20. Dong, L.; Zhao, H.; Zeljkovic, I.; Wilson, S. D.; Harter, J. W. Bulk superconductivity in $FeTe_{1-x}Se_x$ via physicochemical pumping of excess iron. *Phys. Rev. Mater.* **2019**, 3, 114801.

21. Ru, H.; Lin, Y.-S.; Chen, Y.-C.; Feng, Y.; Wang, Y.-H. Observation of two-level critical state in the superconducting FeTe thin films. *Chin. Phys. Lett.* **2019**, 36, 077402.

22. Kawasaki, Y.; Deguchi, K.; Demura, S.; Watanabe, T.; Okazaki, H.; Ozaki, T.; Yamaguchi, T.; Takeya, H.; Takano, Y. Phase diagram and oxygen annealing effect of $FeTe_{1-x}Se_x$ iron-based superconductor. *Solid State Commun.* **2012**, 152, 1135-1138.

23. Si, W.; Jie, Q.; Wu, L.; Zhou, J.; Gu, G.; Johnson, P. D.; Li, Q. Superconductivity in epitaxial thin films of $Fe_{1.08}Te:O_x$. *Phys. Rev. B* **2010**, 81, 092506.

24. Horcas, I.; Fernandez, R.; Gomez-Rodriguez, J. M.; Colchero, J.; Gomez-Herrero, J.; Baro, A. M. WSXM: a software for scanning probe microscopy and a tool for nanotechnology. *Rev. Sci. Instrum.* **2007**, 78, 013705.

25. Zhang, W.-H.; Sun, Y.; Zhang, J.-S.; Li, F.-S.; Guo, M.-H.; Zhao, Y.-F.; Zhang, H.-M.; Peng, J.-P.; Xing, Y.; Wang, H.-C.; Fujita, T.; Hirata, A.; Li, Z.; Ding, H.; Tang, C.-J.; Wang, M.; Wang, Q.-Y.; He, K.; Ji, S.-H.; Chen, X.; Wang, J.-F.; Xia, Z.-C.; Li, L.; Wang, Y.-Y.; Wang, J.; Wang, L.-L.; Chen, M.-W.; Xue, Q.-K.; Ma, X.-C. Direct Observation of high-temperature superconductivity in one-unit-cell FeSe films. *Chin. Phys. Lett.* **2014**, 31, 017401.

26. Liang, J.; Zhang, Y. J.; Yao, X.; Li, H.; Li, Z.-X.; Wang, J.; Chen, Y.; Sou, I. K. Studies on the origin of the interfacial superconductivity of $Sb_2Te_3/Fe_{1+y}Te$ heterostructures. *Pro. Nation. Acad. Sci.* **2020**, 117, 221-227.

27. Telesca, D.; Nie, Y.; Budnick, J. I.; Wells, B. O.; Sinkovic, B. Impact of valence states on the superconductivity of iron telluride and iron selenide films with incorporated oxygen. *Phys. Rev. B* **2012**, 85, 214517.

28. Zhang, Z. T.; Yang, Z. R.; Lu, W. J.; Chen, X. L.; Li, L.; Sun, Y. P.; Xi, C. Y.; Ling, L. S.; Zhang, C. J.; Pi, L.; Tian, M. L.; Zhang, Y. H. Superconductivity in $Fe_{1.05}Te:O_x$ single crystals. *Phys. Rev. B* **2013**, 88, 214511.

29. Smith, N.; Gelting, D.; Basaran, A. C.; Schofield, M.; Schuller, I. K.; Gajdardzijska-Josifovska, M.; Guptasarma, P. Effects of oxygen annealing on single crystal iron telluride. *Physica C Supercond.* **2019**, 567, 1253400.

30. Zhang, Z.; Cai, M.; Li, R.; Meng, F.; Zhang, Q.; Gu, L.; Ye, Z.; Xu, G.; Fu, Y.-S.; Zhang, W. Controllable synthesis and electronic structure characterization of multiple phases of iron telluride thin films. *Phys. Rev. Mater.* **2020**, 4, 125003.

31. Yanpeng Song, T. Y., Xu Chen, Xu Han, Yuan Huang, Xianxin Wu, Andreas P. Schnyder, Jian-Gang Guo, and Xiaolong Chen Enhancement of superconductivity in hole-doped $CsV_3Sb_5$ thin films. *arXiv:2105.09898v1* **2021**.

32. Qn, H.; Guo, B.; Wang, L.; Zhang, M.; Xu, B.; Shi, K.; Pan, T.; Zhou, L.; Chen, J.; Qu, Y.; Xi, B.; Sou, I. K.; Yu, D.; Chen, W.-Q.; He, H.; Ye, F.; Mei, J.-W.; Wang, G. Superconductivity in single-quintuple-layer

Bi$_2$Te$_3$ grown on epitaxial FeTe. *Nano Lett.* **2020**, 20, 3160-3168.

33. Owada, K.; Nakayama, K.; Tsubono, R.; Shigekawa, K.; Sugawara, K.; Takahashi, T.; Sato, T. Electronic structure of a Bi$_2$Te$_3$/FeTe heterostructure: Implications for unconventional superconductivity. *Phys. Rev. B* **2019**, 100, 064518.
34. Guo, B.; Shi, K.-G.; Qin, H.-L.; Zhou, L.; Chen, W.-Q.; Ye, F.; Mei, J.-W.; He, H.-T.; Pan, T.-L.; Wang, G. Evidence for topological superconductivity: Topological edge states in Bi$_2$Te$_3$/FeTe heterostructure. *Chin. Phys. B* **2020**, 29, 097403.
35. Yuan, Y.; Wang, X.; Song, C.; Wang, L.; He, K.; Ma, X.; Yao, H.; Li, W.; Xue, Q.-K. Observation of coulomb gap and enhanced superconducting gap in nano-sized Pb islands grown on SrTiO$_3$. *Chin. Phys. Lett.* **2020**, 37, 017402.

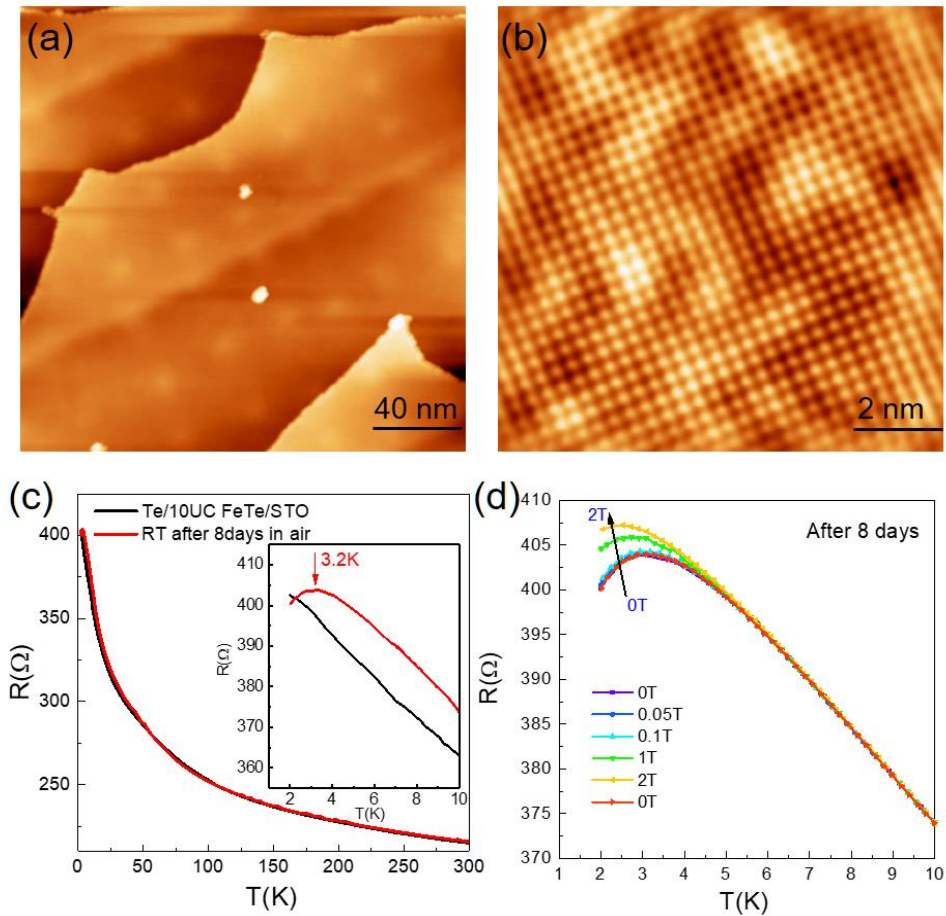

**Figure 1.** (color online) (**a**) Surface topography (200 nm ×200 nm) of the 10UC FeTe films on STO . Scanning condition: sample bias $V_s$=3 V, tunneling current $I_t$=100 pA. (**b**) Typical atomically resolved STM image (10 nm ×10 nm) of 10UC FeTe. Scanning condition: sample bias $V_s$=0.7 V, tunneling current $I_t$=200 pA. (**c**) Temperature dependent resistance (*R-T*) curve of 10UC FeTe thin film on HR-STO before and after exposure to air for 8 days. Inset: *R-T* curve under low temperature, showing turn-down behavior after exposure to air. (**d**) Temperature dependent resistance of 10UC FeTe after exposure to air for 8 days under out-of-plane magnetic field up to 2 T. The turn-down behavior was suppressed under magnetic fields and recovered after magnetic field withdrawal.

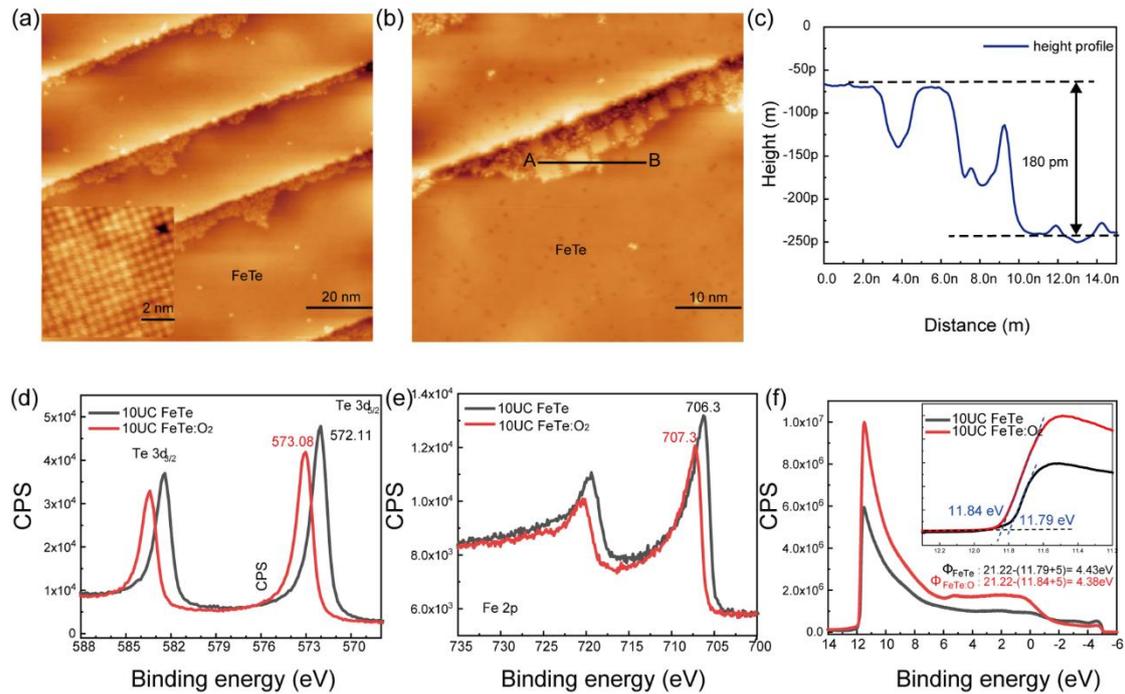

**Figure 2.** (color online) (**a**) Surface topography (100 nm ×100 nm) of the 10UC FeTe films after exposure to oxygen partial pressure of 1.6×10$^{-4}$ mbar for 15 mins at RT. Scanning condition: sample bias $V_s$=1 V, tunneling current $I_t$=100 pA. O$_2$ overlayers were initially formed along the step edge. Inset is the atomically resolved STM image of the uncovered clean FeTe surface. (**b**) Enlarged STM image (50 nm ×50 nm) of O$_2$ overlayers (FeTe:O$_2$) on FeTe surface. Scanning condition: sample bias $V_s$=0.50 V, tunneling current $I_t$=100 pA. (**c**) The height profile of FeTe:O$_2$ along line AB in panel b. (**d, e**) XPS core level spectra of Te 3d (**d**) and Fe 2p (**e**) on as-prepared 10 UC FeTe and FeTe:O$_2$ covered surface, respectively. No trance of Te-O and Fe-O signals can be observed. The spectra show apparent peak shifts after oxygen adsorption. (**f**) UPS spectra of 10UC FeTe covered *w/o* FeTe:O$_2$, and the added voltage here is -5 V. Inset is the large view.

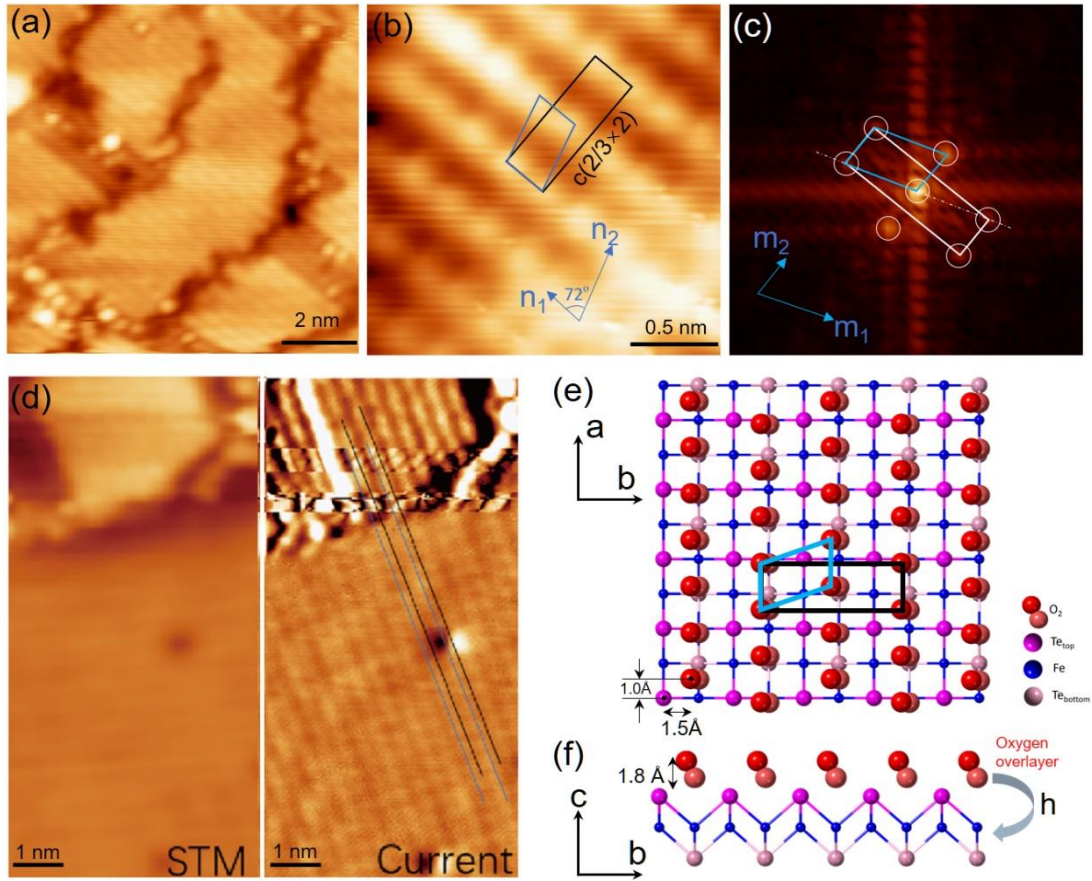

**Figure 3.** (Color online) (**a**) Small compacted FeTe:$O_2$ islands on 10UC FeTe. Dispersed clusters can be observed at domain boundaries. Scanning condition: sample bias $V_s$=1 V, tunneling current $I_t$=200 pA. (**b**) Atomically resolved STM images (2 nm ×2 nm) on FeTe:$O_2$ islands. Scanning condition: sample bias $V_s$=0.1 V, tunneling current $I_t$=200 pA. (**c**) The fast Fourier transform image of (b). White circles mark the reciprocal lattice. (**d**) STM topographic (left panel) and constant current (right panel) images (5 nm × 10 nm) of the 10UC FeTe films with one FeTe:$O_2$ island. The current image helps us to measure the adsorption geometrical configuration. Scanning condition: sample bias $V_s$=0.3 V, tunneling current $I_t$=300 pA. (**e, f**) Schematic diagrams of geometrical configuration of $O_2$ adsorption overlayer on FeTe, (**e**) top side and (**f**) side views.

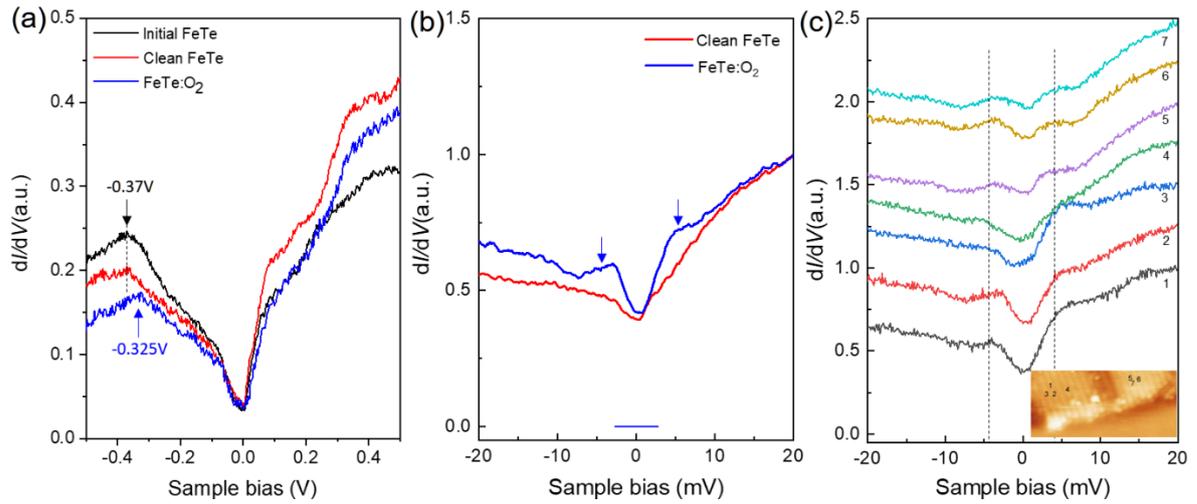

**Figure 4.** (Color online) (**a**) Typical d$I$/d$V$ spectra on 10UC FeTe before and after $O_2$ adsorption. A hump at negative energy shows clear shift on FeTe:$O_2$ overlayer. Starting condition: $V_s$=0.5 V, tunneling current $I_t$=100 pA. (**b**) The zoomed-in d$I$/d$V$ spectra on clean FeTe area and FeTe:$O_2$ overlayer. An energy gap with two pronounced peaks were observed across Fermi level, which is a superconducting gap. Starting condition: $V_s$=20 mV, tunneling current $I_t$=200 pA. (**c**) A series of d$I$/d$V$ spectra on FeTe:$O_2$, showing the energy gap can be observed on the whole FeTe:$O_2$ islands. Starting condition: $V_s$=20 mV, tunneling current $I_t$=200 pA.